\documentstyle[11pt,axodraw]{article}
\begin{document}
\begin{titlepage}
\vspace{-10mm}
\vspace{12pt}
\begin{center}
\begin{large}             
{\bf Classification of constraints\\ using chain by chain method}\\
\end{large}
\vspace{5mm}
{\bf F. Loran 
\footnote{e-mail: farhang@theory.ipm.ac.ir}}
{\bf ,A. Shirzad
\footnote{e-mail:shirzad@cc.iut.ac.ir}}\\
\vspace{12pt} 
{\it Department of  Physics, Isfahan University of Technology \\ 
Isfahan,  IRAN, \\ 
Institute for Studies in Theoretical Physics and Mathematics\\ 
P. O. Box: 5746, Tehran, 19395, IRAN.} 
\vspace{0.3cm}
\end{center}
\abstract{We introduce "chain by chain" method for constructing 
the constraint structure of a system possessing both first and 
second class constraints. We show that the whole constraints 
can be classified into completely irreducible first or second class chains. 
We found appropriate redefinition of second class constraints to obtain 
a symplectic algebra among them.
\ 
\vfill}
\end{titlepage}

\section{Introduction}
Constrained systems, are known from almost 1950 \cite{Dirac,AndBerg}. 
These systems are in fact the basis of gauge 
theories. Modern formalisms of quantization, such as BRST \cite{BRST} and BV \cite{BV}, 
are constructed 
on constrained systems. For a singular Lagrangian system, primary constraints 
emerge since the momenta 
$p_i={\partial L}/{\partial\dot{q}_i}$ are not independent functions of $(q,\dot{q})$. 
However, if we do not care 
about the origin of primary constraints, we can suppose that we are given a canonical Hamiltonian 
$H_c$, together with a number of primary constraints $\phi_1^a,\hspace{1mm}a=1,\ldots,m$.
\par
The equation of motion for an arbitrary function $g(q,p)$ reads \cite{Dirac}:
\begin{eqnarray}\label{i1}
\dot{g}=\{g,H_T\},
\end{eqnarray}
where the total Hamiltonian $H_T$ is defined as
\begin{eqnarray}\label{i2}
H_T=H_c+v_a\phi_1^a,
\end{eqnarray}
in which $v_a$'s are Lagrange multipliers (LM). Since the constraints should vanish at any arbitrary 
time, their consistency from (\ref{i1}) requires that
\begin{eqnarray}\label{i3}
\dot{\phi}^a_1=\{\phi^a_1,H_T\}\approx 0.
\end{eqnarray}
Secondary constraints may emerge from consistency of 
primary constraints. Consistency of any constraint $\chi$ 
implies that $\{\chi,H_T\}\approx 0$, where the weak equality $\approx$ means equality 
on the constraint surface. This may lead to one of the 
following cases \cite{Henbook,Sun}:
\par
$\ \ i$) One of the Lagrange multipliers is determined. Roughly speaking, 
this occurs when $\{\chi,\phi^a_1\}\not\approx 0$ for some $a$.
\par
$\ ii$) Consistency is achieved identically, i.e. $\{\chi,H_c\}\approx 0$ 
and $\{\chi,\phi^a_1\}\approx 0$.
\par
$iii$) A new constraint emerge. This is the case when $\{\chi,\phi^a_1\}\approx 0$ 
but $\{\chi,H_c\}\not\approx o$.
\par
In cases $(i)$ and $(ii)$ the consistency process should stop, but in case (3) 
one should go on through the consistency of 
new constraint $\tilde{\chi}=\{\chi,H_c\}$ and so on. This procedure leads to a {\it constraint 
structure}, which is studied in references such as \cite{Henbook,BatGom,ShiSad}
\par
We assume for simplicity that:
\par
a) No ineffective constraint (such as $\phi=x^2$) emerges.
\par
b) "Bifurcation" does not happen when investigating the consistency conditions, i.e. equations 
like $xp\approx 0$ do not emerge. 
\par
Moreover, we say two functions (in phase space) "commute", when they have vanishing Poisson 
bracket.
\par
Starting with a system of primary constraints $\phi^a_1$, one can use different methods to find 
out all of the constraints.  
The method which is well known in the literature \cite{Henbook,BatGom,HenTeiZan,GoHenPo,GraciaPon} is 
the {\it level by level} method. In this method, at the $n$'th level of consistency, say, 
one tries to solve the equations $\{\phi_n^a,H_T\}\approx 0$ to find the maximum possible 
number of LM's. This depends 
on the rank of matrix $\{\phi^a_n,\phi^b_1\}$. In principle, $\phi^a_n$'s divide into two 
subclasses: $\phi^{a_1}_n$'s which commute with $\phi^a_1$'s and $\phi^{a_2}_n$'s which do not 
commute. The latter i.e. second class constraints, lead to determining a number 
of LM's. The former lead to constraints of the next level, which their number is less 
than or equal to $\phi^{a_1}_n$'s (depending on the number of independent non vanishing functions among 
$\{\phi^{a_1}_n,H_c\}$). As we will discuss in the last section, 
this method and the resulting {\it constraint algebra} is very complicated and a number of its
mathematical statements should begin with "in principle 
it is possible that ...".
\par 
We propose an alternative procedure to produce the constraints of the system. 
Our strategy is as follows:
\par
1) We investigate the consistency of constraints {\it chain by chain}, i.e. 
when a new constraint 
emerge we go on by considering its consistency. In this way, beginning with a 
primary constraint, one produces 
a constraint chain. When a chain terminates, we consider the next primary constraint and 
{\it knit} its corresponding chain and so on. (In some cases we should go back and 
continue to knit 
a chain which had been considered as a terminated one. Such cases will be discussed later.)
 \par
2) All of the equalities should be considered as weak equalities. By weak equality, 
at each step, 
we mean equality on the surface of constraints known up to that step. When a new constraint 
is recognized, it should be added to the set of known constraints. If needed, one can redefine 
any function, such as a constraint or canonical Hamiltonian, by adding a combination of 
known constraints to it.
\par
In this way, the whole system of constraints is managed in a chain structure.  
Moreover, the set of constraints is irreducible by construction.
The details are given in sections 2-4. In sections 2 and 3, we explain the method for 
systems with one or two primary constraints respectively. In section 4 we extend the method to an arbitrary 
system with several primary constraints. Section 5 is devoted to some examples.
\par
The advantages of this method and comparison with the existing level by level method is given in section 6
which is the concluding section.

\section{One Chain System}
Consider a system with one PC, say $\phi_1$. The total Hamiltonian is 
\begin{eqnarray}\label{c1}
H_T=H_c+v\phi_1.
\end{eqnarray}
Following the strategy given in section 1, there is only one chain, 
whose constraints obey the recursion relation 
\begin{eqnarray}\label{c2}
\phi_n=\{\phi_{n-1},H_c\}.
\end{eqnarray}
Suppose first that the chain terminates at level $N$ according to case $(i)$ of 
the previous section, i.e.
\begin{eqnarray}\label{c3}
\{\phi_N,\phi_1\}\approx \eta(q,p)\not\approx 0 
\end{eqnarray}
To investigate the constraint algebra, we arm ourselves with the following lemma. 
\par
{\it Lemma 1:}
\par
For the chain described by (\ref{c2},\ref{c3})
\begin{eqnarray}\label{c5}
\{\phi_j,\phi_i\}&\approx& 0\hspace{1cm} i+j\leq N\\
\label{c6}
\{\phi_{N-i},\phi_{i+1}\}&\approx& (-1)^i\eta\hspace{1cm} i=0,\ldots,N-1
\end{eqnarray}
\par
{\it proof:} Suppose $i<j$. It is obvious that $\{\phi_j,\phi_1\}\approx 0$ for $j<N$, since 
otherwise the chain would be terminated before level $N$. Assuming
\begin{eqnarray}\label{c7}
\{\phi_j,\phi_i\}\approx 0, \hspace{1cm} j=1,\ldots,N-i
\end{eqnarray}
we prove that
\begin{eqnarray}\label{c8}
\{\phi_{j+1},\phi_i\}\approx 0, \hspace{1cm} j=1,\ldots,N-i-1.
\end{eqnarray}
Using (\ref{c2}) and Jacobi identity, we have 
\begin{eqnarray}\label{c9}
\{\phi_j,\phi_{i+1}\}=\{\{\phi_j,\phi_i\},H_c\}-\{\{\phi_j,H_c\},\phi_i\}
\end{eqnarray}
From (\ref{c7}), $\{\phi_j,\phi_i\}$ is a combination of $\phi_k$'s with $k<N$. 
Therefore the first term on RHS of (\ref{c9}) vanishes weakly. The second term 
is just $\{\phi_{j+1},\phi_i\}$ which vanishes according to the assumption (\ref{c7}). 
In this way (\ref{c5}) is proved inductively. The second part of the lemma can be proved 
by inserting $j=N-i$ in (\ref{c9}) which gives
\begin{eqnarray}\label{c10}
\{\phi_{N-i},\phi_{i+1}\}\approx-\{\phi_{N-i+1},\phi_i\}\approx\ldots\approx(-1)^i\{\phi_N,\phi_1\}
\end{eqnarray} 
Then (\ref{c6}) follows from (\ref{c3}). 
\par
Corollaries:
\par
$a)$ All of the constraints of the chain are second class.
\par
$b)$ $N$ is even. (Since in  (\ref{c6}) one should necessarily have $N-i\neq i+1$.)
\par
$c)$ The following diagram shows schematically that each entry in the second half of the chain 
$(\phi_1,\ldots,\phi_K,\phi_{K+1},\ldots,\phi_{2K})$ is {\it conjugate} to (have non vanishing 
Poisson bracket with) its partner in the first half.
\par
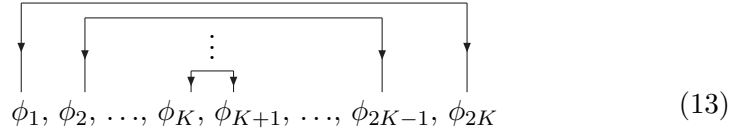
\vspace{1.7cm}
\begin{equation}
\begin{picture}(0,0)(0,0)
\SetScale{0.8}
\ArrowLine(-110,52)(-110,10)
\Line(-110,52)(100,52)
\ArrowLine(100,52)(100,10)
\ArrowLine(-80,45)(-80,10)
\Line(-80,45)(60,45)
\ArrowLine(60,45)(60,10)
\ArrowLine(-30,20)(-30,10)
\Line(-30,20)(-10,20)
\ArrowLine(-10,20)(-10,10)
\Text(-16,28)[]{$\vdots$}
\Text(0,0)[]{$\phi_1$, $\phi_2$, $\ldots$, $\phi_K$, $\phi_{K+1}$, $\ldots$, $\phi_{2K-1}$, $\phi_{2K}$}
\end{picture}
\end{equation}
\par
\vspace{7mm}
We call a chain with the above properties a {\it self-conjugate} (SCC) second class chain.
However, using appropriate redefinitions, one can replace the chain with an equivalent set 
$(\Omega_1,\ldots,\Omega_K,\Omega_{K+1},\ldots,\Omega_N)$ obeying the symplectic algebra:
\begin{equation}\label{in1}
\delta_{ij}=\{\Omega_i,\Omega_j\}\approx J_{ij}
\end{equation} where $J$ is the symplectic $2K\times 2K$ matrix:
\begin{equation}\label{in2}
J=\left(\begin{array}{cc}
\mbox\huge{\bf 0}&\mbox\huge{\bf1}\\
\mbox\huge{\bf -1}&\mbox\huge{\bf 0}\end{array}\right)
\end{equation}
The proof is given in appendix A.
\par
The relation (\ref{in1}) is the best thing that one can find for the algebra of a set of 
second class constraints. In fact, since $\Delta^{-1}=-J$, one can easily define the 
Dirac brackets and get into the reduced phase space.
\par
The other possibility for the chain to terminate at level $N$ is 
$\dot{\phi}_N=\{\phi_N,H_T\}\approx 0$ or equivalently
\begin{eqnarray}\label{c25}
\{\phi_N,\phi_1\}\approx 0\\
\label{c26}
\{\phi_N,H_c\}\approx 0
\end{eqnarray}
Following the steps of the previous lemma one finds that 
\begin{eqnarray}\label{c27}
\{\phi_i,\phi_j\}\approx 0\hspace{1cm} i,j=1,\ldots,N.
\end{eqnarray}
In other words all of the constraints of the chain are first class. We name such a chain 
a first class chain (FCC). 
\par
Concluding, we found that {\it a one-chain system is either completely first class or 
completely second class}. In the former case the Lagrange multiplier remains undetermined 
as an arbitrary function of time ; but in the latter case it would be determined and 
using the Dirac brackets, one can get into the reduced phase space. The following 
flowchart shows the whole procedure schematically.
\par
\pagebreak
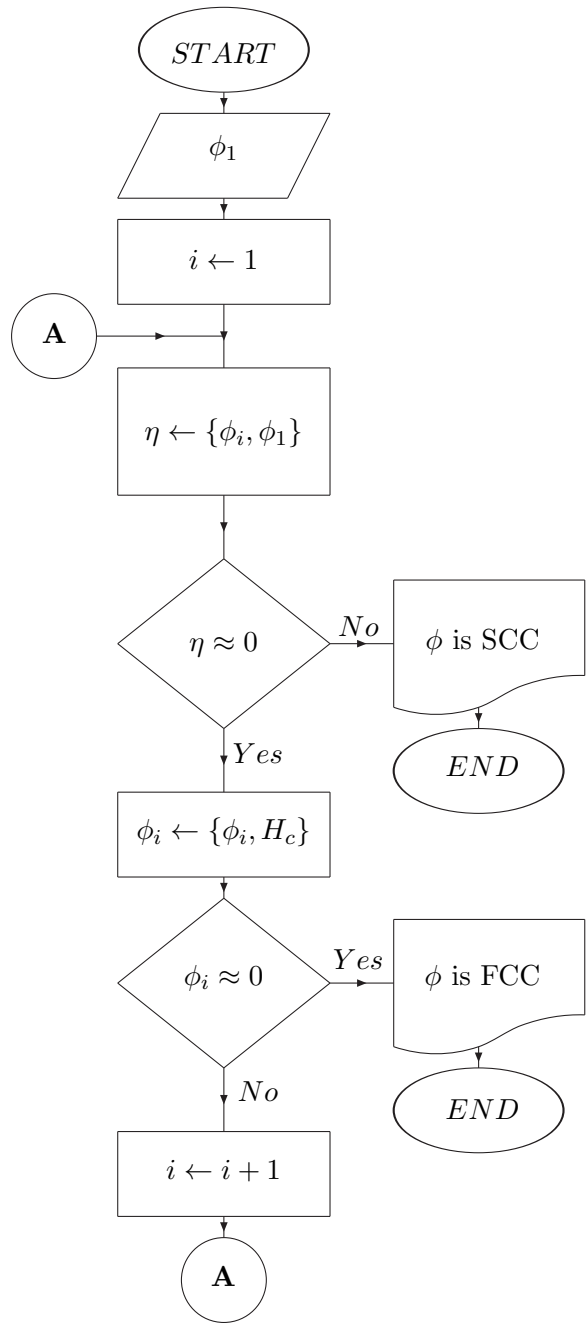
\begin{eqnarray}
\begin{picture}(150,300)(0,0)
\SetScale{0.8}
%
%
\Oval(40,400)(20,40)(0)
\Text(32,320)[]{$START$}
\ArrowLine(40,380)(40,370)
\Line(10,370)(90,370)
\Line(10,370)(-10,330)
\Line(-10,330)(70,330)
\Line(70,330)(90,370)
\Text(32,284)[]{$\phi_1$}
\ArrowLine(40,330)(40,320)
\Line(-10,320)(90,320)
\Line(-10,320)(-10,280)
\Line(-10,280)(90,280)
\Line(90,280)(90,320)
\Text(32,242)[]{$i\leftarrow 1$}
\ArrowLine(40,280)(40,250)
\ArrowLine(-20,265)(40,265)
\Oval(-40,265)(20,20)(0)
\Text(-32,214)[]{{\bf A}}
\Line(-10,250)(90,250)
\Line(-10,250)(-10,190)
\Line(-10,190)(90,190)
\Line(90,190)(90,250)
\Text(32,176)[]{$\eta\leftarrow\{\phi_i,\phi_1\}$}
\ArrowLine(40,190)(40,160)
\Line(40,160)(-10,120)
\Line(-10,120)(40,80)
\Line(40,80)(90,120)
\Line(90,120)(40,160)
\Text(32,96)[]{$\eta\approx 0$}
\ArrowLine(90,120)(120,120)
\Text(83,103)[]{$No$}
\ArrowLine(40,80)(40,50)
\Text(45,55)[]{$Yes$}
%
%
%
\Line(120,150)(120,90)
\Line(120,150)(210,150)
\Line(210,150)(210,105)
\Curve{(210,105)(180,100)(160,90)(120,90)}
\Text(130,97)[]{$\phi$ is SCC}
\ArrowLine(160,90)(160,80)
\Oval(160,60)(20,40)(0)
\Text(130,50)[]{$END$}
\Line(-10,50)(90,50)
\Line(-10,50)(-10,10)
\Line(-10,10)(90,10)
\Line(90,10)(90,50)
\Text(32,25)[]{$\phi_i\leftarrow\{\phi_i,H_c\}$}
\ArrowLine(40,10)(40,0)
\Line(40,0)(-10,-40)
\Line(-10,-40)(40,-80)
\Line(40,-80)(90,-40)
\Line(90,-40)(40,0)
\Text(32,-30)[]{$\phi_i\approx 0$}
\ArrowLine(90,-40)(120,-40)
\Text(83,-23)[]{$Yes$}
\ArrowLine(40,-80)(40,-110)
\Text(45,-73)[]{$No$}
%
%
\Line(120,-10)(120,-70)
\Line(120,-10)(210,-10)
\Line(210,-10)(210,-55)
\Curve{(210,-55)(180,-60)(160,-70)(120,-70)}
\Text(130,-30)[]{$\phi$ is FCC}
\ArrowLine(160,-70)(160,-80)
\Oval(160,-100)(20,40)(0)
\Text(130,-80)[]{$END$}
\Line(-10,-110)(90,-110)
\Line(-10,-110)(-10,-150)
\Line(-10,-150)(90,-150)
\Line(90,-150)(90,-110)
\Text(32,-103)[]{$i\leftarrow i+1$}
\ArrowLine(40,-150)(40,-160)
\Oval(40,-180)(20,20)(0)
\Text(32,-143)[]{{\bf A}}
%
%
%
\end{picture}
\nonumber
\end{eqnarray}
\pagebreak
\section{Two chain system}
Suppose we are given a system with two PC's, say $\phi_1$ and $\psi_1$. 
The total Hamiltonian is
\begin{eqnarray}\label{b1}
H_T=H_c+v\phi_1+w\psi_1.
\end{eqnarray}
Following our strategy, we first go through the consistency of $\phi_1$ and knit the $\phi$-chain via the relation 
\begin{eqnarray}\label{b2}
\phi_{i+1}=\{\phi_i,H_c\}.
\end{eqnarray}
Then we knit the $\psi$-chain in the same way:
\begin{eqnarray}\label{b3}
\psi_{i+1}=\{\psi_i,H_c\}.
\end{eqnarray}
Suppose $\phi_N$ and $\psi_M$ are the last members of the corresponding chains. We define
\begin{eqnarray}\label{b4}
\eta&=&\{\phi_N,\phi_1\}\nonumber\\
\gamma&=&\{\phi_N,\psi_1\}\nonumber\\
\gamma'&=&\{\psi_M,\phi_1\}\nonumber\\
\eta'&=&\{\psi_M,\psi_1\}.
\end{eqnarray}
Our program for constructing the system of constraints are summarized in the 
flowcharts A to E, given below. The function of each part of the program are then briefly 
explained. Lemmas 2-4, which are needed to verify the results will come afterward.

\pagebreak
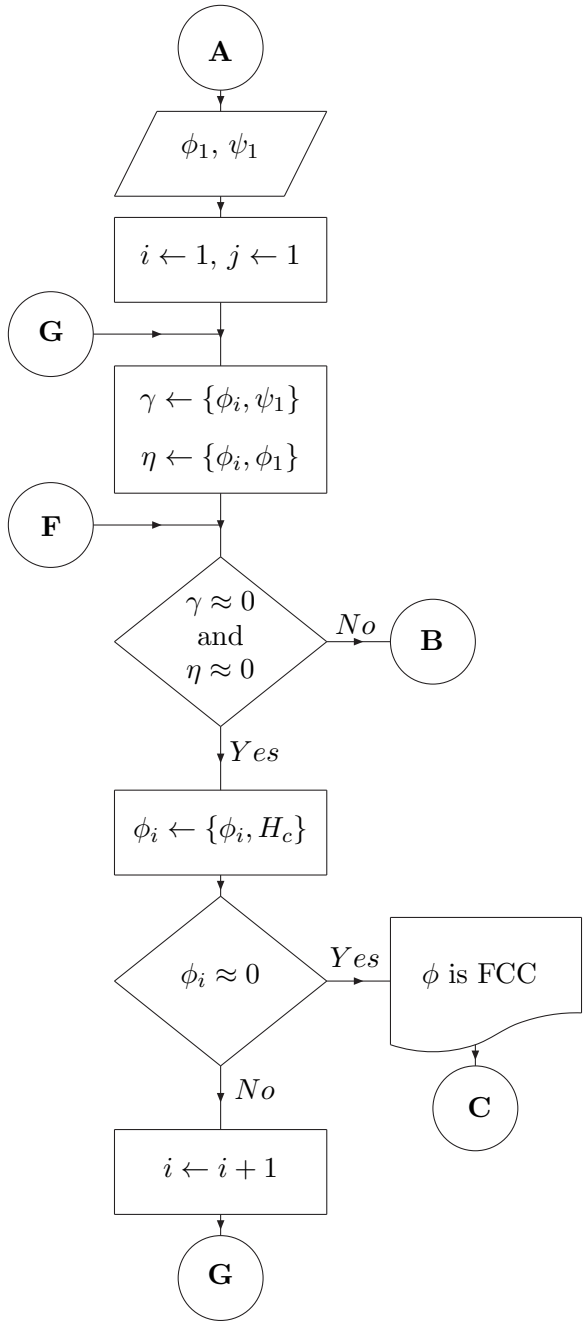
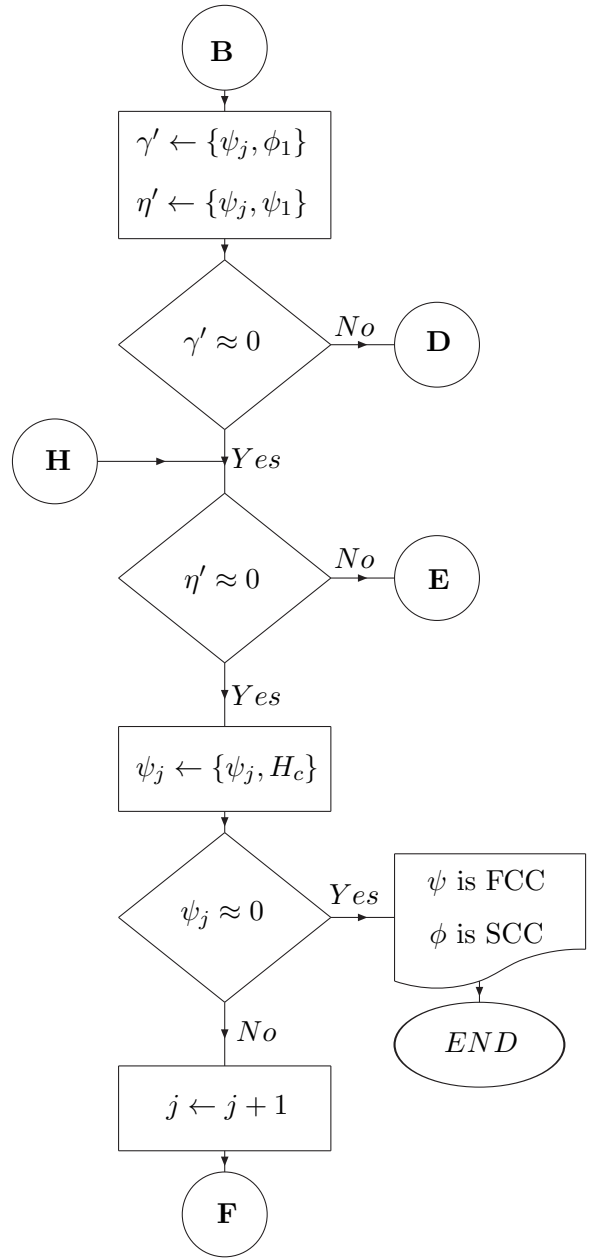
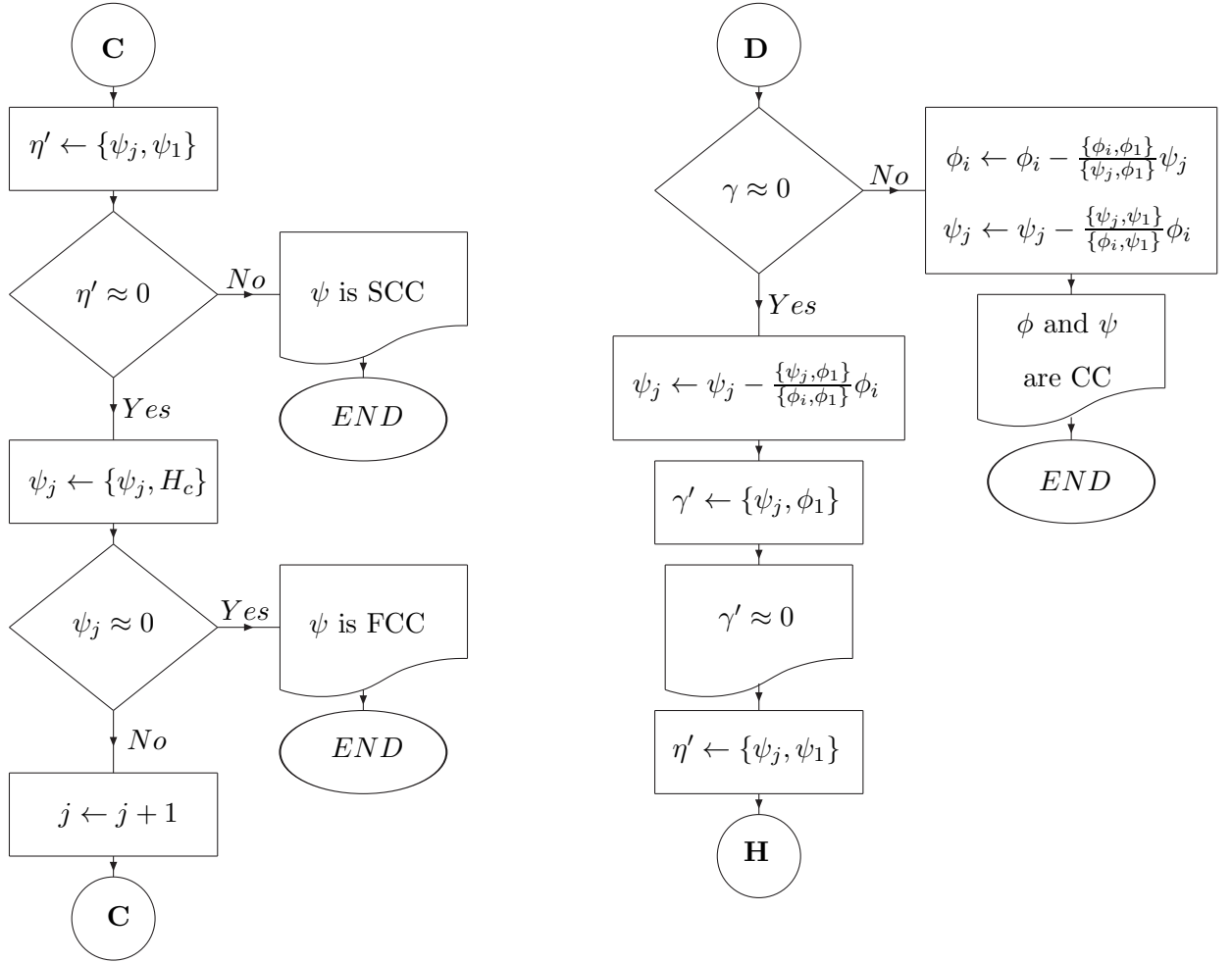
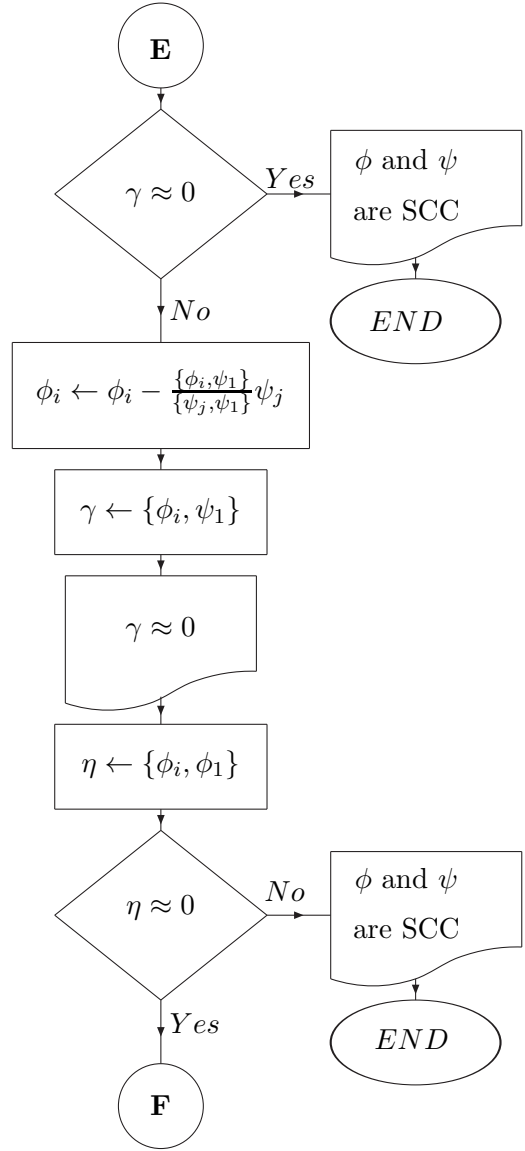
\begin{eqnarray}
\begin{picture}(400,300)(0,0)
\SetScale{0.8}
%
%
\Oval(40,400)(20,20)(0)
\Text(32,320)[]{{\bf A}}
\ArrowLine(40,380)(40,370)
\Line(10,370)(90,370)
\Line(10,370)(-10,330)
\Line(-10,330)(70,330)
\Line(70,330)(90,370)
\Text(32,284)[]{$\phi_1$, $\psi_1$}
\ArrowLine(40,330)(40,320)
\Line(-10,320)(90,320)
\Line(-10,320)(-10,280)
\Line(-10,280)(90,280)
\Line(90,280)(90,320)
\Text(32,242)[]{$i\leftarrow 1$, $j\leftarrow 1$}
\ArrowLine(40,280)(40,250)
\ArrowLine(-20,265)(40,265)
\Oval(-40,265)(20,20)(0)
\Text(-32,214)[]{{\bf G}}
\Line(-10,250)(90,250)
\Line(-10,250)(-10,190)
\Line(-10,190)(90,190)
\Line(90,190)(90,250)
\Text(32,188)[]{$\gamma\leftarrow\{\phi_i,\psi_1\}$}
\Text(32,166)[]{$\eta\leftarrow\{\phi_i,\phi_1\}$}
\ArrowLine(40,190)(40,160)
\ArrowLine(-20,175)(40,175)
\Oval(-40,175)(20,20)(0)
\Text(-32,140)[]{{\bf F}}
\Line(40,160)(-10,120)
\Line(-10,120)(40,80)
\Line(40,80)(90,120)
\Line(90,120)(40,160)
\Text(32,111)[]{$\gamma\approx 0$}
\Text(32,99)[]{and}
\Text(32,85)[]{$\eta\approx 0$}
\ArrowLine(90,120)(120,120)
\Text(83,103)[]{$No$}
\ArrowLine(40,80)(40,50)
\Text(45,55)[]{$Yes$}
\Oval(140,120)(20,20)(0)
\Text(112,97)[]{{\bf B}}
\Line(-10,50)(90,50)
\Line(-10,50)(-10,10)
\Line(-10,10)(90,10)
\Line(90,10)(90,50)
\Text(32,25)[]{$\phi_i\leftarrow\{\phi_i,H_c\}$}
\ArrowLine(40,10)(40,0)
\Line(40,0)(-10,-40)
\Line(-10,-40)(40,-80)
\Line(40,-80)(90,-40)
\Line(90,-40)(40,0)
\Text(32,-30)[]{$\phi_i\approx 0$}
\ArrowLine(90,-40)(120,-40)
\Text(83,-23)[]{$Yes$}
\ArrowLine(40,-80)(40,-110)
\Text(45,-73)[]{$No$}
%
%
\Line(120,-10)(120,-70)
\Line(120,-10)(210,-10)
\Line(210,-10)(210,-55)
\Curve{(210,-55)(180,-60)(160,-70)(120,-70)}
\Text(130,-30)[]{$\phi$ is FCC}
\ArrowLine(160,-70)(160,-80)
\Oval(160,-100)(20,20)(0)
\Text(130,-80)[]{{\bf C}}
\Line(-10,-110)(90,-110)
\Line(-10,-110)(-10,-150)
\Line(-10,-150)(90,-150)
\Line(90,-150)(90,-110)
\Text(32,-103)[]{$i\leftarrow i+1$}
\ArrowLine(40,-150)(40,-160)
\Oval(40,-180)(20,20)(0)
\Text(32,-143)[]{{\bf G}}
%
%
%
%
\Oval(350,400)(20,20)(0)
\Text(280,320)[]{{\bf B}}
\ArrowLine(350,380)(350,370)
\Line(300,370)(400,370)
\Line(300,370)(300,310)
\Line(300,310)(400,310)
\Line(400,310)(400,370)
\Text(280,285)[]{$\gamma'\leftarrow\{\psi_j,\phi_1\}$}
\Text(280,262)[]{$\eta'\leftarrow\{\psi_j,\psi_1\}$}
\ArrowLine(350,310)(350,300)
\Line(350,300)(300,260)
\Line(300,260)(350,220)
\Line(350,220)(400,260)
\Line(400,260)(350,300)
\Text(280,210)[]{$\gamma'\approx 0$}
\ArrowLine(400,260)(430,260)
\Text(330,216)[]{$No$}
\ArrowLine(350,220)(350,190)
\Text(293,166)[]{$Yes$}
\Oval(450,260)(20,20)(0)
\Text(362,210)[]{{\bf D}}
\ArrowLine(290,205)(350,205)
\Oval(270,205)(20,20)(0)
\Text(218,165)[]{{\bf H}}
\Line(350,190)(300,150)
\Line(300,150)(350,110)
\Line(350,110)(400,150)
\Line(400,150)(350,190)
\Text(280,120)[]{$\eta'\approx 0$}
\ArrowLine(400,150)(430,150)
\Text(330,128)[]{$No$}
\ArrowLine(350,110)(350,80)
\Text(293,76)[]{$Yes$}
\Oval(450,150)(20,20)(0)
\Text(362,120)[]{{\bf E}}
\Line(300,80)(400,80)
\Line(300,80)(300,40)
\Line(300,40)(400,40)
\Line(400,40)(400,80)
\Text(282,48)[]{$\psi_j\leftarrow\{\psi_j,H_c\}$}
\ArrowLine(350,40)(350,30)
\Line(350,30)(300,-10)
\Line(300,-10)(350,-50)
\Line(350,-50)(400,-10)
\Line(400,-10)(350,30)
\Text(280,-7)[]{$\psi_j\approx 0$}
\ArrowLine(400,-10)(430,-10)
\Text(330,1)[]{$Yes$}
\ArrowLine(350,-50)(350,-80)
\Text(293,-51)[]{$No$}
%
%
\Line(430,20)(430,-40)
\Line(430,20)(520,20)
\Line(520,20)(520,-25)
\Curve{(520,-25)(490,-30)(470,-40)(430,-40)}
\Text(380,5)[]{$\psi$ is FCC}
\Text(380,-15)[]{$\phi$ is SCC}
\ArrowLine(470,-40)(470,-50)
\Oval(470,-70)(20,40)(0)
\Text(378,-55)[]{$END$}
\Line(300,-80)(400,-80)
\Line(300,-80)(300,-120)
\Line(300,-120)(400,-120)
\Line(400,-120)(400,-80)
\Text(282,-80)[]{$j\leftarrow j+1$}
\ArrowLine(350,-120)(350,-130)
\Oval(350,-150)(20,20)(0)
\Text(282,-120)[]{{\bf F}}
%
%
\end{picture}
\nonumber
\end{eqnarray}

\pagebreak
\begin{eqnarray}
\begin{picture}(400,300)(0,0)
\SetScale{0.8}
%
%
\Oval(40,400)(20,20)(0)
\Text(32,320)[]{{\bf C}}
\ArrowLine(40,380)(40,370)
\Line(-10,370)(90,370)
\Line(-10,370)(-10,330)
\Line(-10,330)(90,330)
\Line(90,330)(90,370)
\Text(32,282)[]{$\eta'\leftarrow\{\psi_j,\psi_1\}$}
\ArrowLine(40,330)(40,320)
\Line(40,320)(-10,280)
\Line(-10,280)(40,240)
\Line(40,240)(90,280)
\Line(90,280)(40,320)
\Text(32,225)[]{$\eta'\approx 0$}
\ArrowLine(90,280)(120,280)
\Text(83,231)[]{$No$}
\ArrowLine(40,240)(40,210)
\Text(45,181)[]{$Yes$}
%
%
\Line(120,310)(120,250)
\Line(120,310)(210,310)
\Line(210,310)(210,265)
\Curve{(210,265)(180,260)(160,250)(120,250)}
\Text(130,225)[]{$\psi$ is SCC}
\ArrowLine(160,250)(160,240)
\Oval(160,220)(20,40)(0)
\Text(130,178)[]{$END$}
\Line(-10,210)(90,210)
\Line(-10,210)(-10,170)
\Line(-10,170)(90,170)
\Line(90,170)(90,210)
\Text(34,152)[]{$\psi_j\leftarrow\{\psi_j,H_c\}$}
\ArrowLine(40,170)(40,160)
\Line(40,160)(-10,120)
\Line(-10,120)(40,80)
\Line(40,80)(90,120)
\Line(90,120)(40,160)
\Text(32,97)[]{$\psi_j\approx 0$}
\ArrowLine(90,120)(120,120)
\Text(83,103)[]{$Yes$}
\ArrowLine(40,80)(40,50)
\Text(45,53)[]{$No$}
%
%
\Line(120,150)(120,90)
\Line(120,150)(210,150)
\Line(210,150)(210,105)
\Curve{(210,105)(180,100)(160,90)(120,90)}
\Text(130,97)[]{$\psi$ is FCC}
\ArrowLine(160,90)(160,80)
\Oval(160,60)(20,40)(0)
\Text(130,50)[]{$END$}
\Line(-10,50)(90,50)
\Line(-10,50)(-10,10)
\Line(-10,10)(90,10)
\Line(90,10)(90,50)
\Text(34,24)[]{$j\leftarrow j+1$}
\ArrowLine(40,10)(40,0)
\Oval(40,-20)(20,20)(0)
\Text(34,-15)[]{{\bf C}}
%
%
%
%
\Oval(350,400)(20,20)(0)
\Text(280,320)[]{{\bf D}}
\ArrowLine(350,380)(350,370)
\Line(350,370)(300,330)
\Line(300,330)(350,290)
\Line(350,290)(400,330)
\Line(400,330)(350,370)
\Text(280,265)[]{$\gamma\approx 0$}
\ArrowLine(400,330)(430,330)
\Text(331,271)[]{$No$}
\ArrowLine(350,290)(350,260)
\Text(294,221)[]{$Yes$}
\Line(430,370)(570,370)
\Line(430,370)(430,290)
\Line(430,290)(570,290)
\Line(570,290)(570,370)
\Text(400,277)[]{$\phi_i\leftarrow\phi_i-\frac{\{\phi_i,\phi_1\}}{\{\psi_j,\phi_1\}}\psi_j$}
\Text(400,250)[]{$\psi_j\leftarrow\psi_j-\frac{\{\psi_j,\psi_1\}}{\{\phi_i,\psi_1\}}\phi_i$}
\ArrowLine(500,290)(500,280)
%
%
%
\Line(455,280)(455,220)
\Line(455,280)(545,280)
\Line(545,280)(545,235)
\Curve{(545,235)(515,230)(495,220)(455,220)}
\Text(400,213)[]{$\phi$ and $\psi$}
\Text(400,193)[]{are CC}
\ArrowLine(500,221)(500,210)
\Oval(500,190)(20,40)(0)
\Text(403,153)[]{$END$}
\Line(280,260)(420,260)
\Line(280,260)(280,210)
\Line(280,210)(420,210)
\Line(420,210)(420,260)
\Text(280,190)[]{$\psi_j\leftarrow\psi_j-\frac{\{\psi_j,\phi_1\}}{\{\phi_i,\phi_1\}}\phi_i$}
\ArrowLine(350,210)(350,200)
\Line(300,200)(400,200)
\Line(300,200)(300,160)
\Line(300,160)(400,160)
\Line(400,160)(400,200)
\Text(280,145)[]{$\gamma'\leftarrow\{\psi_j,\phi_1\}$}
\ArrowLine(350,160)(350,150)
%
%
\Line(305,150)(305,90)
\Line(305,150)(395,150)
\Line(395,150)(395,105)
\Curve{(395,105)(365,100)(345,90)(305,90)}
\Text(280,100)[]{$\gamma'\approx 0$}
\ArrowLine(350,93)(350,80)
\Line(300,80)(400,80)
\Line(300,80)(300,40)
\Line(300,40)(400,40)
\Line(400,40)(400,80)
\Text(280,49)[]{$\eta'\leftarrow\{\psi_j,\psi_1\}$}
\ArrowLine(350,40)(350,30)
\Oval(350,10)(20,20)(0)
\Text(280,10)[]{{\bf H}}
\end{picture}
\nonumber
\end{eqnarray}
\pagebreak

\begin{eqnarray}
\begin{picture}(150,300)(0,0)
\SetScale{0.8}
%
%
\Oval(40,400)(20,20)(0)
\Text(32,320)[]{{\bf E}}
\ArrowLine(40,380)(40,370)
\Line(40,370)(-10,330)
\Line(-10,330)(40,290)
\Line(40,290)(90,330)
\Line(90,330)(40,370)
\Text(32,265)[]{$\gamma\approx 0$}
\ArrowLine(90,330)(120,330)
\Text(81,271)[]{$Yes$}
\ArrowLine(40,290)(40,260)
\Text(43,221)[]{$No$}
%
%
\Line(120,360)(120,300)
\Line(120,360)(210,360)
\Line(210,360)(210,315)
\Curve{(210,315)(180,310)(160,300)(120,300)}
\Text(125,277)[]{$\phi$ and $\psi$}
\Text(125,259)[]{are SCC}
\ArrowLine(160,300)(160,290)
\Oval(160,270)(20,40)(0)
\Text(125,217)[]{$END$}
\Line(-30,260)(110,260)
\Line(-30,260)(-30,210)
\Line(-30,210)(110,210)
\Line(110,210)(110,260)
\Text(32,190)[]{$\phi_i\leftarrow\phi_i-\frac{\{\phi_i,\psi_1\}}{\{\psi_j,\psi_1\}}\psi_j$}
\ArrowLine(40,210)(40,200)
\Line(-10,200)(90,200)
\Line(-10,200)(-10,160)
\Line(-10,160)(90,160)
\Line(90,160)(90,200)
\Text(32,145)[]{$\gamma\leftarrow\{\phi_i,\psi_1\}$}
\ArrowLine(40,160)(40,150)
%
%
\Line(-5,150)(-5,90)
\Line(-5,150)(85,150)
\Line(85,150)(85,105)
\Curve{(85,105)(55,100)(35,90)(-5,90)}
\Text(32,100)[]{$\gamma\approx 0$}
\ArrowLine(40,93)(40,80)
\Line(-10,80)(90,80)
\Line(-10,80)(-10,40)
\Line(-10,40)(90,40)
\Line(90,40)(90,80)
\Text(32,49)[]{$\eta\leftarrow\{\phi_i,\phi_1\}$}
\ArrowLine(40,40)(40,30)
\Line(40,30)(-10,-10)
\Line(-10,-10)(40,-50)
\Line(40,-50)(90,-10)
\Line(90,-10)(40,30)
\Text(32,-5)[]{$\eta\approx 0$}
\ArrowLine(90,-10)(120,-10)
\Text(79,1)[]{$No$}
\ArrowLine(40,-50)(40,-80)
\Text(45,-49)[]{$Yes$}
%
%
\Line(120,20)(120,-40)
\Line(120,20)(210,20)
\Line(210,20)(210,-25)
\Curve{(210,-25)(180,-30)(160,-40)(120,-40)}
\Text(125,5)[]{$\phi$ and $\psi$}
\Text(125,-13)[]{are SCC}
\ArrowLine(160,-40)(160,-50)
\Oval(160,-70)(20,40)(0)
\Text(127,-55)[]{$END$}
\Oval(40,-100)(20,20)(0)
\Text(32,-80)[]{{\bf F}}
\end{picture}
\nonumber
\end{eqnarray}
\par
\vspace{4cm}

{\bf A-} As is apparent from flowchart, this part of program corresponds to 
knitting the $\phi$-chain. The constraint $\phi_i$ may be  the terminating element  
of the chain if $\{\phi_i,\phi_1\}$ and/or $\{\phi_i,\psi_1\}$ does not vanish 
(case $i$ in section 1). If this is the case, one should go to the sub-program 
{\bf B} to  knit the $\psi$-chain. If $\{\psi_i,\phi_1\}$ and $\{\phi_i,\psi_1\}$ 
both vanish, one should consider $\{\phi_i,H_c\}$. The chain terminates 
if $\{\phi_i,H_c\}\approx 0$. Lemma-3 then shows that the $\phi$-chain is first 
class and commute with the $\psi$-chain. The program in this case would go on by knitting 
the $\psi$-chain within the sub-program {\bf C}.
\vspace{3mm}
\par
{\bf B-} Suppose the $\phi$-chain is second class and is already terminated. 
In sub-program {\bf B} we proceed to knitting  the $\psi$-chain afterward. Suppose 
$\{\psi_i,\phi_1\}$ and $\{\psi_i,\psi_1\}$ both vanish. The $\psi$-chain  terminates 
(and is first class) if $\{\psi_i,H_c\}\approx 0$ and it would be continued otherwise. 
Using lemma 3, one can show that conditions $\eta'\approx 0$ 
and $\gamma'\approx 0$ necessarily give $\gamma\approx 0$ and the $\phi$-chain should
have been terminated according to $\eta\not\approx 0$. Therefore, in this case, the $\psi$-chain 
is first class and the $\phi$-chain is self-conjugate. If $\{\psi_j,H_c\}\not\approx 0$ then 
we should continue knitting the 
$\psi$-chain, but we should go first through step {\bf F} which its necessity would be explained below. 
If $\gamma'$ and/or $\eta'$ does not vanish then the $\psi$-chain terminates and 
is second class. Different cases may happen depending on the non-vanishing set 
among $\eta$, $\gamma$, $\eta'$ and $\eta'$. These cases will be considered 
in sub-programs {\bf D} and {\bf E}.
\vspace{3mm}
\par
{\bf C-} Suppose the $\phi$-chain is first class and is terminated. Then the $\psi$-chain 
would be knitted independently. This is done in sub-program {\bf C}, which is almost the 
same as what we did for one chain system. The $\psi$-chain may be first or second class 
and will be terminated in the usual manner in each case. 
\vspace{3mm}
\par
{\bf D-} If $\gamma'\not\approx 0$ and $\gamma\not\approx 0$ (right branch in the 
flowchart {\bf D}) then one can redefine $\phi_N$ and $\psi_M$ to insure that $\eta\approx 0$ 
and $\eta' \approx 0$. So we have two second class chains in such a way that the 
terminating element of each chain does not commute with the top element of the other 
chain. We call this second class system a {\it cross-conjugate} (CC) two-chain system. 
In lemma 4 we show that the chains have the same length and their elements are pairwise 
conjugate. If $\gamma'\not\approx 0$ but $\gamma\approx 0$ (necessarily $\eta\not\approx 0$) 
then using the redefinition shown in the flowchart we make $\gamma'$ to vanish and then 
come back to stage {\bf H} of the sub-program {\bf B}. 
\vspace{3mm}
\par
{\bf E-} Suppose $\gamma'\approx 0$ and $\eta'\not\approx 0$. If $\gamma \approx 0$, 
then necessarily $\eta\not\approx 0$ and we have two self-conjugate second class chains. 
It will be shown in lemma 4 that by suitably redefining $H_c$ the chains would commute 
(i.e. $\{\phi_i,\psi_j\}\approx 0$ for each $\phi_i$ and $\psi_j$). 
If $\gamma\not\approx 0$ one can make it to vanish by redefining $\phi_i$ as indicated in 
the flowchart. This alteration may change $\eta$; if it still does not vanish, 
the situation is like above and we have two self-conjugate chains. 
If $\eta$ vanishes as a consequence of redefining $\phi_i$, one should come back to the stage 
{\bf F} of the sub-program {\bf A} and continue knitting the $\phi$-chain.
\vspace{3mm}
\par
{\bf F-} An essential point is that during knitting the $\psi$-chain, it may happen 
that the previously non-vanishing functions $\eta$ or $\gamma$ do vanish on the surface 
of new constraints. If so, one should come back to $\phi$-chain and continue knitting it. 
Therefore, when a new constraint emerges in $\psi$-chain we should be sure 
that $\eta$ and $\gamma$ are still non vanishing. If at least one of them is remained nonzero 
then we are allowed to go to the beginning point of sub-program {\bf B} and go on.
\par
As is seen, after running the program we may have one of the following possibilities for the two chain system:
\par
\ \ $i$) two first class chains,
\par
\ $ii$) one first class and one self-conjugate chain,
\par     
$iii$) two self-conjugate chains,
\par
\ $iv$) a system of two cross-conjugate chains.
\par
In cases $i$-$iii$ the constraints of each chain commute with the constraints  of the other 
chain, but in case $iv$ each constraint in one chain is conjugate to its partner in the 
other chain.
\par
Now let us give some details that we encountered in the program as three lemmas.
\par
{\it lemma 2:} If the  $\phi$-chain is FCC (relation (\ref{c27})), then $\{\phi_i,\psi_j\}\approx 0$ 
for $i=1,\ldots,N$ and $j=1,\ldots,M$.
\par
{\it proof:} Since the $\phi$-chain is first class, it is obvious that 
$\{\phi_i,\psi_1\}\approx 0$ for $i=1,\ldots,N$. Asuming $\{\phi_i,\psi_j\}\approx 0$  
for a given $\psi_j$ and all $\phi_i$, we prove that $\{\phi_i,\psi_{j+1}\}\approx 0$. 
To do this, note that
\begin{eqnarray}\label{b5}
\{\phi_i,\psi_{j+1}\}=\{\{\phi_i,\psi_j\},H_c\}-\{\phi_{i+1},\psi_j\}
\end{eqnarray}
where we have used the Jacobi identity and recursion relations (\ref{b2}) and (\ref{b3}). 
The second term on the RHS of (\ref{b5}) vanishes weakly according to our assumption. 
(In case  
$i=N$, one should consider a combination of $\phi_i$'s instead of $\phi_{N+1}$.) 
In the first term on RHS of (\ref{b5}) $\{\phi_i,\psi_j\}$ is a combination of constraints 
and has weakly vanishing Poisson bracket with $H_c$. The only exemption is when $j=M$ 
and the $\psi$-chain is second class. In this case one can redefine $H_c$ as follows: 
\begin{eqnarray}\label{b6}
H_c\to H_c-\frac{\{\psi_M,H_c\}}{\{\psi_M,\psi_1\}}\psi_1
\end{eqnarray}
This redefinition ensures us that $\{\psi_M,H_c\}\approx 0$ and the lemma holds always.
\par
{\it lemma 3:} If $\gamma\approx 0$,$\gamma'\approx 0$,$\eta\not\approx 0$ 
and $\eta'\not\approx 0$, then the chains can be made to commute.
\par
{\it proof:} Redefining $H_c$ as
\begin{eqnarray}\label{b8}
H_c\to H_c-\frac{\{\phi_N,H_c\}}{\{\phi_N,\phi_1\}}\phi_1-
\frac{\{\psi_M,H_c\}}{\{\psi_M,\psi_1\}}\psi_1
\end{eqnarray}
we have $\{\phi_N,H_c\}\approx 0$ and $\{\psi_M,H_c\}\approx 0$. Then following the 
same steps as we did in lemma 2, this lemma will also be proved.
\par
{\it lemma 4:} If $\gamma\not\approx 0$, $\gamma'\not\approx 0$, $\eta\approx 0$ and 
$\eta'\approx 0$, then the chains have the same length and their constraints are pairwise 
conjugate to each other.
\par
{\it proof:} Using Jacobi identity and recursion relations (\ref{b2}) and (\ref{b3}) 
it is easy to show that
\begin{eqnarray}\label{b7}
\{\phi_N,\psi_1\}\approx -\{\phi_{N-1},\psi_2\}\approx\ldots\approx (-1)^N\{\phi_1,\psi_N\}
\end{eqnarray}
This relation shows that the $\psi$-chain really terminates after $N$ steps (i.e. $M=N$) 
and the conjugate pairs are as follows:$\left(\phi_N,\psi_1\right),
\left(\phi_{N-1},\psi_2\right),\ldots,\left(\phi_1,\psi_N\right)$.
\par
\section{Multi-chain System}
Suppose we are given $m$ primary constraints $\phi^a_1,\hspace{1mm} a=1,\ldots,m$. The 
process of constructing the constraint chains is roughly a generalization of what we did
for two-chain systems. However, one should first arrange the primary constraints 
in such a way that the set of primary second class constraints come first as cross-conjugate pairs.
Suppose we have knitted constraint chains up to some particular chain 
which begins with $\phi_1^a$. We want to show how the typical chain $\phi^a$ is produced. We 
assume that $\phi^a_1$ commute with all previous constraints; otherwise the chain $\phi^a$ 
should be conjugate to one of the previous chains. The following flowchart shows the main 
process of constructing the chain $\phi^a$.
\par
\pagebreak
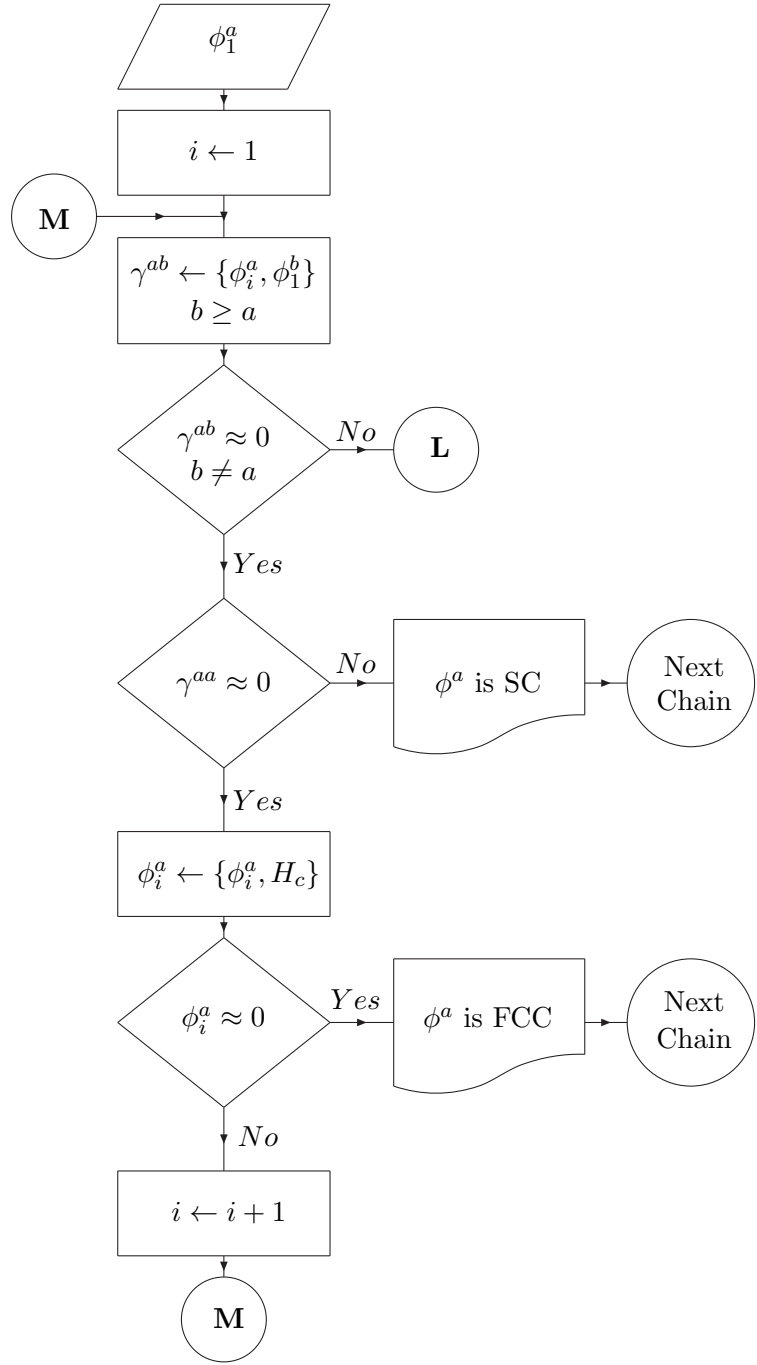
\begin{eqnarray}
\begin{picture}(150,350)(0,0)
\SetScale{0.8}
\Line(10,470)(90,470)
\Line(10,470)(-10,430)
\Line(-10,430)(70,430)
\Line(70,430)(90,470)
\Text(32,364)[]{$\phi_1^a$}
\ArrowLine(40,430)(40,420)
\Line(-10,420)(90,420)
\Line(-10,420)(-10,380)
\Line(-10,380)(90,380)
\Line(90,380)(90,420)
\Text(32,322)[]{$i\leftarrow 1$}
\ArrowLine(40,380)(40,360)
\ArrowLine(-20,370)(40,370)
\Oval(-40,370)(20,20)(0)
\Text(-32,296)[]{{\bf M}}
\Line(-10,360)(90,360)
\Line(-10,360)(-10,310)
\Line(-10,310)(90,310)
\Line(90,310)(90,360)
\Text(32,276)[]{$\gamma^{ab}\leftarrow\{\phi_i^a,\phi_1^b\}$}
\Text(32,261)[]{$b\ge a$}
\ArrowLine(40,310)(40,300)
\Line(40,300)(-10,260)
\Line(-10,260)(40,220)
\Line(40,220)(90,260)
\Line(90,260)(40,300)
\Text(32,214)[]{$\gamma^{ab}\approx 0$}
\Text(32,200)[]{$b\ne a$}
\ArrowLine(90,260)(120,260)
\Text(82,216)[]{$No$}
\ArrowLine(40,220)(40,190)
\Text(45,166)[]{$Yes$}
\Oval(140,260)(20,20)(0)
\Text(114,210)[]{{\bf L}}
\Line(40,190)(-10,150)
\Line(-10,150)(40,110)
\Line(40,110)(90,150)
\Line(90,150)(40,190)
\Text(32,120)[]{$\gamma^{aa}\approx 0$}
\ArrowLine(90,150)(120,150)
\Text(82,128)[]{$No$}
\ArrowLine(40,110)(40,80)
\Text(45,76)[]{$Yes$}
%
%
%
\Line(120,180)(120,120)
\Line(120,180)(210,180)
\Line(210,180)(210,135)
\Curve{(210,135)(180,130)(160,120)(120,120)}
\Text(132,120)[]{$\phi^a$ is SC}
\ArrowLine(210,150)(230,150)
\Oval(260,150)(30,30)(0)
\Text(210,127)[]{Next}
\Text(210,114)[]{Chain}
\Line(-10,80)(90,80)
\Line(-10,80)(-10,40)
\Line(-10,40)(90,40)
\Line(90,40)(90,80)
\Text(34,48)[]{$\phi_i^a\leftarrow\{\phi_i^a,H_c\}$}
\ArrowLine(40,40)(40,30)
\Line(40,30)(-10,-10)
\Line(-10,-10)(40,-50)
\Line(40,-50)(90,-10)
\Line(90,-10)(40,30)
\Text(32,-7)[]{$\phi_i^a\approx 0$}
\ArrowLine(90,-10)(120,-10)
\Text(82,1)[]{$Yes$}
\ArrowLine(40,-50)(40,-80)
\Text(45,-51)[]{$No$}
%
%
\Line(120,20)(120,-40)
\Line(120,20)(210,20)
\Line(210,20)(210,-25)
\Curve{(210,-25)(180,-30)(160,-40)(120,-40)}
\Text(132,-7)[]{$\phi^a$ is FCC}
\ArrowLine(210,-10)(230,-10)
\Oval(260,-10)(30,30)(0)
\Text(210,0)[]{Next}
\Text(210,-14)[]{Chain}
\Line(-10,-80)(90,-80)
\Line(-10,-80)(-10,-120)
\Line(-10,-120)(90,-120)
\Line(90,-120)(90,-80)
\Text(34,-80)[]{$i\leftarrow i+1$}
\ArrowLine(40,-120)(40,-130)
\Oval(40,-150)(20,20)(0)
\Text(34,-120)[]{{\bf M}}

\end{picture}
\nonumber
\end{eqnarray}
\par
\pagebreak

In this process (except at step {\bf M} which will be explained) we have nothing to do with
the previous chains. This means that to treating with some constraint $\phi^a_1$, it is sufficient 
to consider its Poisson brackets with $\phi^b_1$'s for $b\ge a$. There are some small 
differences with two-chain systems at steps {\bf L} and {\bf M} which will be explained below.
\vspace{3mm}
\par
{\bf L-} Suppose $\gamma^{ab}$ defined as $\{\phi^a_1,\phi^b_1\}$ does not vanish for some $b\ne a$. 
One can move $\phi^b_1$ to the position next after $\phi^a_1$ and knit the corresponding 
chain. The system of chains $\phi^a$ and $\phi^b$ is a two-chain system, which should be 
analyzed in the same way as mentioned in the previous section. If more than one primary 
constraint do not commute with $\phi^a_1$, then one can choose one of them 
and redefine the other 
ones in a way similar to appendix A to make them commute with $\phi^a_1$.
\vspace{3mm}
\par
{\bf M-} Similar to the step {\bf F} of the two-chain system, when a new constraint $\phi^a_i$ 
emerges, one should check whether the previously assumed non vanishing functions like 
$\eta$'s, $\gamma$'s, etc, (which appear 
in the end points of second class chains) are still non vanishing or not. For example, 
if we have had a
self-conjugate chain $\phi^c$, i.e. $\gamma^{cc}=\{\phi^c_{N_c},\phi^c_1\}$ have been 
non vanishing and it vanishes after the emergence of a new constraint one should come back 
to the chain $\phi^c$ and continue to knit it.
\par
After all, the set of constraints would be classified as a number of first class chains, 
self-conjugate second class chains and some pairs of cross-conjugate second class chains. 
As it is shown through the lemmas, it is easy to make the chains to commute with each other 
(only chains in a pair of cross-conjugate chains do not commute). 
Using the methods given in the appendices A and B, 
it is also possible, to redefine the second class chains 
such that their Poisson brackets obey the symplectic algebra. 
\section{Examples}
\ \ {it i}) Consider the Lagrangian
\begin{equation}
L=\dot{x}\dot{z}+\frac{1}{2}\dot{\alpha}^2+xy+\alpha\beta.
\label{e1}\end{equation}
The primary constraints are 
\begin{equation}
\phi_1=p_y\hspace{5mm},\hspace{5mm}\psi_1=p_\beta, 
\label{e2}\end{equation}
and the canonical Hamiltonian can be written as
\begin{equation}
H_c=p_xp_z+\frac{1}{2}p_\alpha ^2-xy-\alpha\beta 
\label{e4}\end{equation}
By knitting the chains, we have two constraint chains as follows
\begin{equation}
\begin{array}{lllll}
\phi_1=p_y&&&\psi_1=p_\beta\\
\phi_2=x&&&\psi_2=\alpha\\
\phi_3=p_z&&&\psi_3=p_\alpha\\
&&&\psi_4=\beta
\end{array}
\label{e5}\end{equation}
Clearly, the first chain is FC and the second one is self-conjugate.
\vspace{3mm}
\par
\ {\it ii}) As an example of cross-conjugate chains, consider the system given by
\begin{equation}
L=\dot{x}\dot{z}+\dot{\alpha}\dot{\beta}+xy+\alpha\gamma+z\gamma.
\label{e7}\end{equation}
Primary constraints are 
\begin{equation}
\phi_1=p_y\hspace{5mm},\hspace{5mm}\phi_2=p_\gamma
\label{e8}\end{equation}
and the canonical Hamiltonian is 
\begin{equation}
H_c=p_xp_z+p_\alpha p_\beta -xy-\alpha\gamma-z\gamma.
\label{e10}\end{equation}
It is obvious that the chains 
\begin{equation}
\begin{array}{lllll}
\phi_1=p_y&&&\psi_1=p_\gamma\\
\phi_2=x&&&\psi_2=z+\alpha\\
\phi_3=p_z&&&\psi_3=p_x+p_\beta\\
\phi_4=\gamma &&&\psi_4=y
\end{array}
\label{e11}\end{equation}
are cross-conjugate. 
\vspace{3mm}
\par
{\it iii}) As an example of multi chain system, consider the QED Lagrangian:
\begin{eqnarray}\label{qe1}
L=\int d^3x \left[-\frac{1}{4}F_{\mu\nu}F^{\mu\nu}+\overline{\psi}\gamma^{\mu}(i\partial_{\mu}-eA_{\mu})\psi-m\overline{\psi}\psi\right].
\end{eqnarray}
Primary constraints and the canonical Hamiltonian are as follows:
\begin{eqnarray}\label{qe2}
\phi^1_1=\Pi^0,\hspace{5mm}\phi^2_1=\overline{\Pi},\hspace{5mm}\phi^3_1=\Pi-i\overline{\psi}\gamma^0,
\end{eqnarray}
\begin{eqnarray}\label{qe3}
H_c=\int d^3x\left[\frac{1}{2}\Pi_i^2+A_0\partial_i\Pi_i+\frac{1}{4}F_{ij}F_{ij}-\overline{\psi}i\gamma^i\partial_i\psi-e\overline{\psi}\gamma^\mu A_\mu\psi+m\overline{\psi}\psi\right],
\end{eqnarray}
where $\Pi^\mu$, $\Pi$ and $\overline{\Pi}$ are momenta conjugate to $A^\mu$, $\psi$ and $\overline{\psi}$ respectively. 
\par
Since $\{\phi^2_1,\phi^3_1\}=i\gamma^0$, these constraints form a cross conjugate pair. Consequently, according to 
lemma 3, we should redefine the canonical 
Hamiltonian $H_c$ (relation \ref{b8}),
\begin{eqnarray}
H'_c=H_c-\phi^3_1\frac{1}{\{\phi^2_1,\phi^3_1\}}\{\phi^2_1,H_c\}-\{\phi^3_1,H_c\}\frac{1}{\{\phi^3_1,\phi^2_1\}}\phi^2_1.
\nonumber
\end{eqnarray}
As mentioned before, this is necessary in order that the remaining chain 
commute with $\phi^2_1$ and $\phi^3_1$. Consequently, the $\phi^1$-chain arises as 
\begin{eqnarray}
\phi^1_1&=&\Pi^0\nonumber\\
\phi^1_2&=&-\partial_i\Pi_i-ie(\Pi\psi-\overline{\psi}\,\overline{\Pi}).
\end{eqnarray}
The interesting point is that without redefining $H_c$ the $\phi^1$-chain would terminate at $\phi'^1_2=-\partial_i\Pi_i
+e\overline{\psi}\gamma^0\psi$. This constraint does not commute with $\phi^2_1$ and $\phi^3_1$. In the literature 
\cite{Sun} the problem is solved by searching for a first class combination of $\phi^2_1$, $\phi^3_1$ and $\phi'^1_2$. As 
we see, this difficulty is removed automatically in our method.
\section{Conclusion}
Given the primary constraints, the main task, in a constrained system, is to find the 
set of secondary constraints in such a way that the consistency of all constraints is achieved 
on the constraint surface, i.e. all the constraints commute weakly with $H_c$. The well known 
level by level method, explained in section 1, has some characteristics as follows:
\par
1) At a given level, one should separate first and second class constraints from each other. 
This goal can be achieved in simple examples possessing a few number of constraints. 
However, for general cases, no method is introduced. 
 
\par
2) There is not a simple algebraic relation between the constraints of two adjacent 
levels, since finding the independent functions among $\{\phi^{a_1}_n,H_c\}$ (see the 
introduction) needs algebraic manipulations, which in the general case may combine these 
functions with each other.
\par
3) It may happen that the rank of matrix $\{\phi^a_n,\phi^b_1\}$ reduces at subsequent levels. 
In other words, some of the previously assumed second class constraints may appear to be first 
class when the constraint surface extends due to emergence of new constraints. This important 
point, though considered in the literature 
\cite{GoHenPo,GraciaPon}, but most of the time is bypassed by 
the assumption that "the rank of the matrix of Poisson brackets of the constraints is constant during consistency process."
\par
4) In general, the constraint algebra is not so simple. In other words, it is difficult 
to find simple algebraic relations between Poisson brackets of constraints. 
In fact, due to this difficulty, most studies on constrained systems 
have been done for special cases such as pure first class or pure second class systems.
\par
We think that in chain by chain method, most of the above difficulties can be removed or reduced. 
In this method, one encounters some systematic operations (summarized as 
flowcharts) which can be followed easily.
The second class constraints, in this method, are definitely separated from 
first class ones 
in a natural way, which makes their algebra very simple. In this way, most of the results,  
achieved through literature for purely first class systems or purely second class 
systems, can be easily generalized for arbitrary systems possessing both first and second class 
constraints.
\par
As mentioned in introduction, the chain structure and irreducibility are the main results of our method. 
One important application is in constructing the generating function of gauge transformations, 
which is a well known problem in the context of gauge theories \cite{ShiSha}. For example, the authors of \cite{Chai93} have 
tried to prove the existence of the gauge generator, but they encountered some difficulties since the constraints 
in their chain structure are not necessarily irreducible. In some cases, such as for Abelian constraints, the chain 
structure makes it easy to solve the conditions that the gauge generator should satisfy. However, in our opinion, the 
chain by chain method is able to provide a new framework for discussing the problem of constructing the gauge generator 
for the general cases.
\par
As another important application, the chain by chain method can be better applied in the gauge 
fixing process \cite{ShiLo}. In that paper, we have shown that one should introduce only gauge fixing 
constraints which are conjugate to the last elements of the chains. Then, their consistency would 
give us the other necessary gauge fixing constraints. As is apparent, such a simple method for 
fixing the gauges can not be found in the context of level by level method. 
\vspace{5mm} 

{\bf Acknowledgements}

We would like to thank Prof. A. A. Deriglazov for his useful comment. 
\pagebreak

\vspace{3mm}
{\large{\bf Appendix A}}
\vspace{3mm}

Here we prove that for a 1-chain system the constraints can be put in a way so that their algebra has the form given 
by (\ref{in1},\ref{in2}). 
Let us first rename the constraints in the second half of the chain such that
\begin{equation}
\phi_i^*=(-1)^i\phi_{N-i+1}\hspace{1cm}i=1,\ldots,K.
\label{ap1}\end{equation}
In this way lemma 1 reads 
\begin{eqnarray}\label{ap2}
\{\phi_i,\phi_j\}&\approx& 0\hspace{1cm}i,j=1,\ldots,K\\
\label{ap3}
\{\phi_i,\phi_j^*\}&\approx& 0\hspace{1cm}i=1,\ \ \ \ldots,K,\ j>i\\
\label{ap4}
\{\phi_i,\phi_i^*\}&\approx& \eta.
\end{eqnarray}
Now we can redefine the set $(\phi_1,\ldots,\phi_K;\phi_1^*,\ldots,\phi^*_K)$ such that
\begin{eqnarray}\label{ap5}
{\tilde\phi_1}&=&\phi_1,\nonumber\\
{\tilde\phi^*_1}&=&\phi_1^*,
\end{eqnarray}
\begin{eqnarray}\label{ap6}
{\tilde\phi_i}&=&\phi_i-\sum^{i-1}_{k=1}\frac{\{\phi_i,{\tilde\phi_k^*}\}}{\{{\tilde\phi_k},{\tilde\phi_k^*}\}}{\tilde\phi_k},
\end{eqnarray}
\begin{equation}\label{ap7}
{\tilde\phi_i^*}=\phi_i^*-\sum^{i-1}_{k=1}\frac{\{\phi_i^*,{\tilde\phi_k^*}\}}{\{{\tilde\phi_k},{\tilde\phi_k^*}\}}{\tilde\phi_k}.
\end{equation}
Using (\ref{ap2}) it is easy to show that 
\begin{equation}\label{ap8}
\{{\tilde\phi_i},{\tilde\phi_j}\}\approx 0.
\end{equation}
We can also show that
\begin{equation}\label{ap9}
\{{\tilde\phi_i},{\tilde\phi^*_i}\}\approx \delta_{ij}\eta.
\end{equation}
To prove this assertion, consider a definite $j$. For $i<j$, using (\ref{ap7}) and (\ref{ap8})
 it is obvious that 
$\{{\tilde\phi_i},{\tilde\phi_j^*}\}\approx 0$. For $i=j$ 
using (\ref{ap6}-\ref{ap8}) and (\ref{ap4}) one can write
\begin{eqnarray}\label{ap10}
\{{\tilde\phi_i},{\tilde\phi_i^*}\}&\approx &\{\phi_i,\phi_i^*\}\nonumber\\&\approx &\eta
\end{eqnarray}
For $i>j$ we prove (\ref{ap9}) inductively. First one can see that (\ref{ap9}) is true for $i=2$ and $j=1$, i.e.
\begin{equation}\label{ap11}
\{{\tilde\phi_2},{\tilde\phi_1^*}\}=\left\{\phi_2-\frac{\{\phi_2,\phi_1\}}{\{\phi_1,\phi_N\}}\phi_1\ ,\ \phi_N\right\}\approx 0,
\end{equation}
where we have used (\ref{ap5}),(\ref{ap6}) and (\ref{ap1}). Suppose that (\ref{ap9}) holds at all steps up to a definite step $i$. This means that
\begin{equation}
\{{\tilde\phi_k},{\tilde\phi_j^*}\}\approx 0\hspace{1cm}k\ne j,\ k=2,\ldots,i.
\label{ap12}\end{equation}
Then we show that (\ref{ap12}) is also true for $k=i+1$. For this reason, consider
\begin{equation}
\{{\tilde\phi_{i+1}},{\tilde\phi_j^*}\}=\left\{\phi_{i+1}-
\sum^i_{k=1}\frac{\{\phi_{i+1},{\tilde\phi_k^*}\}}{\{{\tilde\phi_k},{\tilde\phi_k^*}\}}{\tilde\phi_k}\ ,\ {\tilde\phi_j^*}\right\}.
\label{ap13}\end{equation}
Using (\ref{ap12}) only the term $k=j$ remains in the sum over $k$, which would be cancelled by construction, to give
\begin{equation}
\{{\tilde\phi_{i+1}},{\tilde\phi_j^*}\}\approx 0.
\label{ap14}\end{equation}
Finally, we prove that 
\begin{equation}\label{ap15}
\{{\tilde\phi^*_i},{\tilde\phi^*_j}\}\approx 0.
\end{equation}
Suppose $i>j$. Using (\ref{ap6}) for ${\tilde\phi_i^*}$ and then (\ref{ap10}), 
the proof is straightforward. Putting (\ref{ap8}), (\ref{ap9}) and (\ref{ap15}) altogether, 
the desired algebra (\ref{in1}) of the text would be obtained by defining
\begin{equation}
\begin{array}{l}
\Omega_i=\eta ^{-1}{\tilde\phi_i}\vspace{2mm}\\
\Omega_{K+i}={\tilde\phi_i^*}
\end{array}\hspace{1cm}i=1,\ldots,K
\label{ap16}\end{equation}

\vspace{3mm}
{\large{\bf Appendix B}}
\vspace{3mm}

In this appendix, we generalize the results of appendix A. 
Consider the two cross-conjugate chains
\begin{equation}
\begin{array}{cccc}
\phi_1&&&\psi_1\\
\vdots&&&\vdots\\
\phi_N&&&\psi_N
\end{array}
\label{bp1}\end{equation}
with algebra
\begin{equation}
\{\phi_N,\psi_1\}\approx (-1)^N\{\psi_N,\phi_1\}\approx\eta
\label{bp2}\end{equation}
If $N$ is even one can replace $\phi_1$ and $\psi_1$ with  $\xi_1=\phi_1+\psi_1$ and $\zeta_1=\phi_1-\psi_1$. 
This would result to the chains
\begin{equation}
\begin{array}{cccc}
\xi_1=\phi_1+\psi_1&&&\zeta_1=\phi_1-\psi_1\\
\vdots&&&\vdots\\
\xi_1=\phi_N+\psi_N&&&\zeta_N=\phi_N-\psi_N
\end{array}
\label{bp3}\end{equation}
In this way the algebra of the constraints is changed to 
\begin{equation}
\begin{array}{l}
\{\xi_N,\zeta_1\}\approx \{\zeta_N,\xi_1\}\approx 0\\
\{\xi_N,\xi_1\}\approx\{\zeta_N,\zeta_1\}\approx2\eta.
\end{array}
\label{bp4}\end{equation}
As is observed, we have replaced a pair of cross-conjugate chains with two self-conjugate ones. The remainder of 
the procedure is as in the case of one chain system. 
That is, following the steps given in appendix A, one can reach to a symplectic algebra among the constraints. 
As an example, the reader can test the Lagrangian (\ref{e7}) with chains given in (\ref{e11}).
\par
Now suppose $N$ is odd. 
If $\{\phi_N,H_c\}\approx 0$, from lemma 3, the elements of the $\phi$-chain commute 
with each other. Noticing (\ref{b7}) one can see that by defining 
$\phi^*_i$ as
\begin{equation}
\phi^*_i=(-1)^i\psi_{N-i+1}\hspace{5mm}i=1,\ldots,N
\label{bp5}\end{equation}
the same algebra of (\ref{ap2})-(\ref{ap4}) will be reproduced. Therefore, one can follow the procedure of appendix A to reach the desired goal.
If $\{\phi_N,H_c\}\not \approx 0$, but instead $\{\psi_N,H_c\}\approx 0$, the same thing can be done, this time with redefining $\phi_i$'s as
\begin{equation}
\psi^*_i=(-1)^i\phi_{N-i+1}.
\label{bp6}\end{equation}
The only considerable case occurs when
\begin{equation}
\begin{array}{l}
\{\phi_N,H_c\}=\gamma\not\approx 0\\
\{\psi_N,H_c\}=\chi\not\approx 0
\end{array}
\label{bp7}\end{equation}
This time we consider
\begin{equation}
\xi_1=\chi\phi_1-\gamma\psi_1
\label{bp8}\end{equation}
as the primary constraint of the first chain. After $N-1$ levels of consistency, one would obtain
\begin{equation}
\xi_N=\chi\phi_N-\gamma\psi_N
\label{bp9}\end{equation}
such that 
\begin{equation}
\{\xi_N,H_c\}\approx 0
\label{bp10}\end{equation}
as can be seen directly from (\ref{bp7}). Moreover, using (\ref{bp2}) one can see
\begin{equation}
\{\xi_N,\xi_1\}\approx\chi\gamma(1+(-1)^N)\eta
\label{bp11}\end{equation}
which vanishes for $N$ odd.
Again the elements of the $\xi$-chain commute with each other, so we continue like the 
previous case.

\end{document}